\documentclass[11pt, letterpaper]{article}

\usepackage[letterpaper,left=1in,right=1in,top=1in,bottom=1in]{geometry}
\usepackage{graphicx}
\usepackage{epstopdf}
\usepackage{amsmath}
\usepackage{amssymb}
\usepackage{mathrsfs}
\usepackage{setspace}
\usepackage{epstopdf}
\usepackage{color}
\usepackage{amsmath,amssymb,amsfonts}
\usepackage{enumitem}
\usepackage{booktabs}
\usepackage{multirow}
\usepackage{cite}
\usepackage{siunitx,etoolbox} 
\usepackage{arydshln}
\usepackage{subfigure}
\usepackage{changes}
\usepackage{makecell}
\newtheorem{rmk}{Remark}

\newtheorem{lemma}{Lemma}
\newtheorem{definition}{Definition}

\usepackage[ruled,linesnumbered, vlined]{algorithm2e}

\usepackage[definethebibliography]{easybib}
\graphicspath{ {202406-Deep DeePC/figures/} }

\title{\Large \bf Deep DeePC: Data-enabled predictive control with low or no online optimization using deep learning}

\author{
\centerline{\normalsize Xuewen Zhang$^{a}$, Kaixiang Zhang$^{b}$, Zhaojian Li$^{b}$, Xunyuan Yin$^{a,c,}$\thanks{Corresponding author: X. Yin. Tel: (+65) 6316 8746. Email: xunyuan.yin@ntu.edu.sg.}}
\vspace{5mm}\\
\centerline{\small $^{a}$School of Chemistry, Chemical Engineering and Biotechnology, Nanyang Technological University,}\\
\centerline{\small 62 Nanyang Drive, 637459, Singapore} \\
\centerline{\small $^{b}$Department of Mechanical Engineering, Michigan State University, East Lansing, MI 48824, USA}\\
\centerline{\small $^{c}$ Nanyang Environment and Water Research Institute (NEWRI), Nanyang Technological University,}\\
\centerline{\small 1 CleanTech Loop, 637141, Singapore}\\
}
\allowdisplaybreaks

\date{}

\begin{document}

\maketitle
\thispagestyle{plain}
\pagestyle{plain}
\setstretch{1.5}

\begin{abstract}

Data-enabled predictive control (DeePC) is a data-driven control algorithm that utilizes data matrices to form a non-parametric representation of the underlying system, predicting future behaviors and generating optimal control actions. DeePC typically requires solving an online optimization problem, the complexity of which is heavily influenced by the amount of data used, potentially leading to expensive online computation. In this paper, we leverage deep learning to propose a highly computationally efficient DeePC approach for general nonlinear processes, referred to as Deep DeePC. Specifically, a deep neural network is employed to learn the DeePC vector operator, which is an essential component of the non-parametric representation of DeePC. This neural network is trained offline using historical open-loop input and output data of the nonlinear process. With the trained neural network, the Deep DeePC framework is formed for online control implementation. At each sampling instant, this neural network directly outputs the DeePC operator, eliminating the need for online optimization as conventional DeePC. The optimal control action is obtained based on the DeePC operator updated by the trained neural network. To address constrained scenarios, a constraint handling scheme is further proposed and integrated with the Deep DeePC to handle hard constraints during online implementation. The efficacy and superiority of the proposed Deep DeePC approach are demonstrated using two benchmark process examples.

\end{abstract}

\section*{Introduction}

Model predictive control (MPC) has been widely used for advanced process control of nonlinear industrial processes~\cite{findeisen2002introduction, mayne2000constrained}. Developing a nonlinear MPC typically requires a high-fidelity first-principles model that accurately describes the dynamic behavior of the underlying process~\cite{graber2012nonlinear, palma2022integration, findeisen2004computational}.
Complex industrial processes are increasingly utilized across various fields to enhance operational efficiency, production consistency, and product quality~\cite{daoutidis2018integrating, christofides2013distributed, daoutidis2016sustainability}. The growing scale and structural complexity of these processes present significant challenges in developing nonlinear first-principles dynamic process models. Data-based optimal control offers a promising alternative for developing advanced control approaches for nonlinear processes without the need for an accurate first-principles model~\cite{rajulapati2022integration, markovsky2021behavioral, lamnabhi2017systems, tang2018distributed, tang2022data, hu2023machine, wang2024integrating}.

The Koopman theory-based framework~\cite{koopman1931hamiltonian} holds great promise for data-driven modeling and control of nonlinear systems. By constructing a linear model within a lifted state-space for a general nonlinear system from offline data, Koopman modeling facilitates the application of linear control theory to nonlinear systems and processes. Notable advancements include Koopman-based model predictive control (MPC) approaches~\cite{korda2018linear, narasingam2019koopman, han2023robust, han2020deep, zhang2023reduced}, which develop linear MPC schemes for nonlinear systems, thereby maintaining convex online optimization despite the inherent nonlinearity of the underlying systems and processes. Koopman-based modeling and control have been widely applied across various fields, including industrial processes~\cite{narasingam2020application, li2024data, shi2023data, li2024machine}, vehicles and robotics~\cite{shi2021enhancement, cibulka2020model, xiao2022deep}, and power systems~\cite{netto2018robust, zhao2023deep, xu2023data}. The development of Koopman MPC typically requires full-state feedback from the systems, with a few exceptions~\cite{han2024efficient, de2024koopman}. An input-output Koopman control method was proposed, where the Koopman-based controller only requires measurements of partial state variables directly used for calculating operational costs~\cite{han2024efficient}. A non-exact multi-step Koopman predictor was identified using only input and output data for the predictive control of nonlinear systems~\cite{de2024koopman}. We note that, in practical applications, measuring certain state variables can be challenging or costly. The need for full-state measurements poses significant challenges to the broader applications of Koopman-based control methods in complex industrial systems.

In recent years, data-enabled predictive control (DeePC) has received increasing attention for its ability to address constrained optimal control using only input and output data~\cite{zhang2022data, coulson2019data, huang2023robust}. DeePC leverages Willems' fundamental lemma~\cite{willems2005note} to form a non-parametric representation of the system using a Hankel matrix established based on pre-collected input and output trajectories. Willems' fundamental lemma indicates that any input and output trajectories of a linear system are spanned by the Hankel matrix when the pre-collected input trajectory is persistently exciting~\cite{willems2005note}. DeePC produces the optimal control action by solving an optimization problem in a receding-horizon manner, which is similar to MPC~\cite{garcia1989model}. In the optimization problem, the number of the decision variables is dependent on the dimension of the Hankel matrix. To satisfy the persistently exciting condition, a sufficiently large amount of historical data is typically needed. This can lead to relatively large dimensions for both the Hankel matrix and the decision variables, which will increase the computational complexity of the online optimization problem associated with DeePC.

In the existing literature, efforts have been focused on addressing the computational challenges of DeePC by reducing the size or complexity of the Hankel matrix and optimization problems. Singular value decomposition was leveraged on collected data to reduce the dimensionality of the DeePC problem, thus decreasing computational complexity~\cite{zhang2023dimension}. Proper orthogonal decomposition was implemented to formulate a reduced order control problem, which was subsequently solved using a customized MPC solver~\cite{carlet2022real}. LQ factorization was employed to redefine the decision variables, replacing the Hankel matrix with a lower triangular matrix~\cite{sader2023causality, breschi2023data}. Lagrange multiplier was used to obtain a nominal solution to the DeePC optimization problem, and an adaption approach was proposed to efficiently update the nominal solution without recomputing the optimal solution~\cite{vahidi2023data}. The primal-dual algorithm was leveraged to solve the DeePC problem recursively, and the fast Fourier transform was used to expedite the computation of Hankel matrix-vector products~\cite{baros2022online}. A size-invariant differentiable convex problem was proposed to learn the scoring function of the DeePC problem recursively, where the scoring function evaluates the likelihood of the predicted input and output trajectories belonging to the system~\cite{zhou2024learning}. While these existing methods improve the computational feasibility of DeePC, they still involve solving optimization problems or iterating algorithms during real-time applications, highlighting ongoing challenges in achieving real-time efficiency.

Motivated by these observations, this study aims to integrate deep learning methods with DeePC to eliminate the need to solve the online optimization problem at every sampling instant during the online implementation of a DeePC-based predictive controller. By leveraging a deep neural network, we learn the vector operator of DeePC from historical data collected from open-loop operations. The trained neural network outputs the vector operator without solving an optimization at each sampling instant during online implementation. Additionally, we develop a constraint handling scheme and integrate this scheme with the proposed optimization-free DeePC to address constrained case scenarios. The proposed method is further slightly extended to address cases where reference inputs are unavailable. The proposed method is evaluated through two case studies on a gene regulatory network and a chemical process. 

The contributions of this work and the main highlights of the proposed approach are summarized as follows:
\begin{enumerate}[label=(\alph*), leftmargin=2\parindent]
    \item We propose a deep learning-enabled DeePC method, referred to as Deep DeePC, to improve online computational efficiency; this method does not need to solve any online optimization when constraints are either absent or not violated.
    \item A constraint handling scheme is developed and integrated with Deep DeePC to realize constrained optimal control while maintaining a low optimization load.
    \item The Deep DeePC control scheme can be trained using historical input and output data collected from open-loop process operations.
    \item The proposed approach can be implemented even when reference inputs corresponding to the desired reference outputs are not given.
    \item We use two benchmark examples to evaluate the performance of the proposed approach. Good control performance is achieved, system constraints are satisfied, and the computation time is significantly reduced as compared to conventional DeePC.
\end{enumerate}

\section*{Preliminaries and problem formulation}

\subsection*{Notation}

$\mathbb{R}$ denotes the set of real numbers. $\mathbb{Z}_{>0}$ and $\mathbb{Z}_{\geq 0}$ denote the sets of positive integers and the set of non-negative integers, respectively. $\mathbb{E}$ denotes the expectation. $\Vert x\Vert_Q^2$ is the square of the weighted Euclidean norm of vector $x$ with positive-definite weighting matrix $Q$, computed as $\Vert x\Vert_Q^2:=x^\top Q x$. $\text{diag}(\cdot)$ denotes a diagonal matrix. $I_n$ is an identity matrix of dimension $n$. $\odot$ denotes the Hadamard product. $\mathbf{1}_{a<b}$ denotes an indicator function; it equals $1$ if $a<b$, and equals $0$ otherwise. $\mathcal{U}(a, b)$ denotes a uniform distribution with lower and upper bounds $a$ and $b$, respectively. $x(i)$ denotes the $i$th variable of state vector $x$. $x_k \in \mathbb{R}^{n_x}$ is the state vector at time instant $k$, and $\{x\}_{j}^{l}:=[x_j^\top, \ldots, x_l^\top]^\top$ contains the state sequence from time instant $j$ to $l$. 
 
\subsection*{Non-parametric representation of linear systems}

Consider a discrete linear time-invariant (LTI) system in which the dynamic behaviors are described by the following state-space form:
\begin{equation}\label{lti-sys}
    \begin{aligned}
        x_{k+1} &= A x_k + B u_k \\
        y_k &= C x_k + D u_k
    \end{aligned}
\end{equation}
where $x_k \in \mathbb{X} \subset \mathbb{R}^{n_x}$ is the system state vector; $u_k \in \mathbb{U} \subset \mathbb{R}^{n_u}$ is the control input vector; $y_k \in \mathbb{Y} \subset \mathbb{R}^{n_y}$ is the system output vector; $A \in \mathbb{R}^{n_x \times n_x}$, $B \in \mathbb{R}^{n_x \times n_u}$, $C \in \mathbb{R}^{n_y \times n_x}$, and $D \in \mathbb{R}^{n_y \times n_u}$ are system matrices; $\mathbb{X}$, $\mathbb{U}$, and $\mathbb{Y}$ are compact sets.

Let $L$, $T \in \mathbb{Z}_{> 0}$ and $T \geq L$. Let $\mathbf{u}_T^d := \{u^d\}_1^T \in \mathbb{R}^{n_u T}$ and  $\mathbf{y}_T^d := \{y^d\}_1^T \in \mathbb{R}^{n_y T}$ denote the input and output sequences for $T$ time instants, respectively. The superscript $d$ indicates the corresponding data are historical data. For the input sequence $\mathbf{u}_T^d$, the Hankel matrix of depth $L$ is defined as follows:
\begin{equation}\label{hankel}
    \mathscr{H}_L(\mathbf{u}_T^d) :=\left[\begin{array}{c c c c }
                    u_1^d & u_2^d & \cdots & u_{T-L+1}^d \\
                    u_2^d & u_3^d & \cdots & u_{T-L+2}^d \\
                    \vdots & \vdots &\ddots &\vdots \\
                    u_L^d & u_{L+1}^d & \cdots & u_{T}^d
                    \end{array}\right]
\end{equation}
where $\mathscr{H}_L(\mathbf{u}_T^d) \in \mathbb{R}^{n_u L \times (T-L+1)}$. Accordingly, for the controlled output sequence $\mathbf{y}_T^d$, define the Hankel matrix $\mathscr{H}_L(\mathbf{y}_T^d) \in \mathbb{R}^{n_y L \times (T-L+1)}$. Next, we introduce the concept of persistent excitation and Willems' fundamental lemma~\cite{willems2005note}.

\begin{definition}\label{def1}
Let $T$, $L \in \mathbb{Z}_{> 0}$ and $T \geq L$. The input sequence $\mathbf{u}_T := \{u\}_1^T$ is persistently exciting of order $L$, if $\mathscr{H}_L(\mathbf{u}_T)$ is of full row rank.
\end{definition}

\begin{lemma}\label{lemma1}(Willems' fundamental lemma~\cite{willems2005note})
Consider an LTI system (\ref{lti-sys}) and assume this system is controllable. 
Consider that $\mathbf{u}_T^d := \{u^d\}_1^T \in \mathbb{R}^{n_u T}$ and  $\mathbf{y}_T^d := \{y^d\}_1^T \in \mathbb{R}^{n_y T}$ are the T-step input and output sequences for system (\ref{lti-sys}), respectively, and the input sequence $\mathbf{u}_T^d$ is persistently exciting of order $L+n_x$. Any $L$-step sequences $\mathbf{u}_L :=\{u\}_1^L\in \mathbb{R}^{n_u L} $ and $\mathbf{y}_L :=\{y\}_1^L\in \mathbb{R}^{n_y L}$ are the input and output trajectories of system (\ref{lti-sys}), if and only if 
\begin{equation}\label{wfl}
    \left[\begin{array}{c}
            \mathscr{H}_L(\mathbf{u}_T^d) \\
            \mathscr{H}_L(\mathbf{y}_T^d)
          \end{array}
    \right] g = 
    \left[\begin{array}{c}
         \mathbf{u}_L \\
         \mathbf{y}_L 
    \end{array}
    \right]
\end{equation}
for vector $g \in \mathbb{R}^{T-L+1}$.
\end{lemma}

A persistently exciting input sequence should be sufficiently rich and of sufficient length to excite the system. This produces an output sequence that is sufficient to represent the behavior of the underlying system. Based on Lemma \ref{lemma1}, a non-parametric representation of the system (\ref{lti-sys}) can be formulated using finite input and output sequences when the input sequence is persistently exciting.

\subsection*{Data-enabled predictive control (DeePC)}

Data-enabled predictive control (DeePC)~\cite{coulson2019data} is a data-based control approach that creates a non-parametric representation of the underlying system using pre-collected input and output data. This way, system identification/modeling can be bypassed.

Lemma \ref{lemma1} allows for the description of the dynamic behaviors of the system using data collected offline. Let $T_{ini}$, $N_p \in \mathbb{Z}_{>0}$ and $L=T_{ini}+N_p$. Any $L$-step input and output sequence of the system (\ref{lti-sys}) can be expressed using pre-collected data. To conduct an $N_p$-step prediction, the Hankel matrices $\mathscr{H}_L(\mathbf{u}_T^d)$ and $\mathscr{H}_L(\mathbf{y}_T^d)$ are partitioned into two parts, that is, the past data of length $T_{ini}$ and the future data of length $N_p$, described as follows:
\begin{equation}\label{past-future}
    \left[ \begin{array}{c}
                U_p \\
                U_f
           \end{array}
    \right] := \mathscr{H}_L(\mathbf{u}_T^d), \quad
    \left[ \begin{array}{c}
                Y_p \\
                Y_f
           \end{array}
    \right] := \mathscr{H}_L(\mathbf{y}_T^d)
\end{equation}
where $U_p \in \mathbb{R}^{n_u T_{ini} \times (T-L+1)}$ denotes the past data which consist of the first $n_u T_{ini}$ block rows of $\mathscr{H}_L(\mathbf{u}_T^d)$; $U_f \in \mathbb{R}^{n_u N_p \times (T-L+1)}$ denotes the future data which consist of the last $n_u N_p$ block rows of $\mathscr{H}_L(\mathbf{u}_T^d)$ (similarly for $Y_p$ and $Y_f$).

In online implementation, at time instant $k$, let $\mathbf{u}_{ini,k} :=\{u\}_{k-T_{ini}}^{k-1}$ and $\mathbf{y}_{ini,k} :=\{y\}_{k-T_{ini}}^{k-1}$ be the $T_{ini}$-step input and output sequences before the current time instant $k$, respectively. Let $\hat{\mathbf{u}}_k :=\{\hat{u}\}_{k|k}^{k+N_p-1|k}$ and $\hat{\mathbf{y}}_k :=\{\hat{y}\}_{k|k}^{k+N_p-1|k}$ be the future $N_p$-step input and output prediction, where $\hat{u}_{j|k}$ and $\hat{y}_{j|k}$ represent the predicted input and output for time instant $j$ obtained at time instant $k$, respectively. Based on Willems' fundamental lemma, $\mathbf{u}_{ini,k}$, $\mathbf{y}_{ini,k}$, $\hat{\mathbf{u}}_k$, and $\hat{\mathbf{y}}_k$ sequences belong to the system (\ref{lti-sys}), if and only if there exists a vector $g_k$ at time instant $k$ such that:
\begin{equation}\label{wfl-pf}
    \left[\begin{array}{c}
            U_p \\
            Y_p \\
            U_f \\
            Y_f
          \end{array}
    \right] g_k = 
    \left[\begin{array}{c}
         \mathbf{u}_{ini,k} \\
         \mathbf{y}_{ini,k} \\
         \hat{\mathbf{u}}_k \\
         \hat{\mathbf{y}}_k
    \end{array}
    \right]
\end{equation}
The vector $g_k$ in (\ref{wfl-pf}) is referred to as ``DeePC operator" in the remainder of this paper. The initial input and output trajectories, $\mathbf{u}_{ini,k}$ and $\mathbf{y}_{ini,k}$, determine the initial state $x_k$ of the underlying system, which is the starting point for the future trajectories~\cite{coulson2019data, markovsky2008data}.

Based on (\ref{wfl-pf}), the optimization problem associated with DeePC at time instant $k$ can be formulated as follows:
\begin{subequations}\label{deepc}
    \begin{align}
        \min_{g_k, \hat{\mathbf{u}}_k, \hat{\mathbf{y}}_k} \  &\Vert \hat{\mathbf{y}}_k - \mathbf{y}^r_k \Vert_Q^2 + \Vert \hat{\mathbf{u}}_k -\mathbf{u}^r_k\Vert_R^2\label{deepc:1} \\
        \label{deepc:2}\text{s.t.} \quad &\left[\begin{array}{c}
            U_p \\
            Y_p \\
            U_f \\
            Y_f
          \end{array}
    \right] g_k = 
    \left[\begin{array}{c}
         \mathbf{u}_{ini,k} \\
         \mathbf{y}_{ini,k} \\
         \hat{\mathbf{u}}_k \\
         \hat{\mathbf{y}}_k
    \end{array} \right]  \\
    &\hat{u}_{j|k} \in \mathbb{U}, \quad j = k, \ldots, k+N_p-1 \\
    &\hat{y}_{j|k} \in \mathbb{Y}, \quad j = k, \ldots, k+N_p-1
    \end{align} 
\end{subequations}
where $\mathbf{y}^r_k := \{y^r\}_{k}^{k+N_p-1} \in \mathbb{R}^{n_y N_p}$ and $\mathbf{u}^r_k := \{u^r\}_{k}^{k+N_p-1} \in \mathbb{R}^{n_u N_p}$ are the reference trajectories of controlled output and control input, respectively; $Q \in \mathbb{R}^{n_y N_p \times n_y N_p}$ and $R \in \mathbb{R}^{n_u N_p \times n_u N_p}$ are tunable weighting matrices.

In the online implementation, DeePC solves (\ref{deepc}) in a receding horizon manner~\cite{coulson2019data, zhang2023dimension}. Specifically, at sampling instant $k$, $k \in \mathbb{Z}_{\geq 0}$, the optimal DeePC operator $g^*_k$, optimal control input sequence $\hat{\mathbf{u}}_k^*$, and optimal predicted output trajectory $\hat{\mathbf{y}}_k^*$ are obtained by solving (\ref{deepc}), and the first control action $\hat{u}^*_{k|k}$ in the optimal control sequence $\hat{\mathbf{u}}^*_k = [\hat{u}_{k|k}^{* \ \top}, \ldots, \hat{u}_{k+N_p-1|k}^{* \ \top}]^\top$ will be applied to the system to achieve desired control performance. At the next sampling instant $k+1$, $\mathbf{u}_{ini,k+1}$ and $\mathbf{y}_{ini,k+1}$ are updated with the applied input and measured output data, $u_{k}$ and $y_{k}$, from the previous time instant $k$, respectively.

\subsection*{Problem formulation}
In this work, we consider discrete-time nonlinear systems of which the dynamics can be described by the following state-space form:
\begin{equation}\label{eq:nonlinear sys}
\begin{aligned}
            x_{k+1} &= f(x_k, u_k) \\
            y_k &= C x_k
\end{aligned}
\end{equation}
where $x_k \in \mathbb{X} \subset \mathbb{R}^{n_x}$ is the system state vector; $u_k \in \mathbb{U} \subset \mathbb{R}^{n_u}$ is the control input vector; $y_k \in \mathbb{Y} \subset \mathbb{R}^{n_y}$ is the output vector; $f: \mathbb{X} \times \mathbb{U} \to \mathbb{X}$ is a nonlinear function that characterizes the state dynamics of the process; $C \in \mathbb{R}^{n_y \times n_x}$ is the system matrix; $\mathbb{X}$, $\mathbb{U}$, and $\mathbb{Y}$ are compact sets. We consider nonlinear systems in which the controlled outputs linearly depend on the state vectors. Our objective is to address the control for the nonlinear processes using only input and output data in an efficient manner within the DeePC framework.

Similar to model predictive control, a conventional DeePC-based controller needs to solve online optimization in a receding horizon manner at every new sampling instant~\cite{coulson2019data, zhang2023dimension}. The optimization problem associated with DeePC can be computationally complex, although DeePC typically formulates convex optimization~\cite{zhang2023dimension, carlet2022real, zhou2024learning}. Based on this consideration, in this work, we aim to propose an efficient DeePC method that requires low optimization/no optimization during online implementation.

To achieve this objective, we propose to use a deep neural network to approximate the DeePC operator $g$ of the non-parametric formulation in (\ref{wfl-pf}). Historical open-loop data are used for the training of the deep neural network, and are formulated into initial and reference trajectories following the DeePC framework. Once the training is completed, the neural network will output the values of the DeePC operator $g$ at each sampling instant without solving the optimization in the form of (\ref{deepc}).

Additionally, to handle the cases when the control action generated by the deep neural network leads to constraint violations, we propose a constraint handling scheme to ensure that the input and output constraints of the system are maintained during online implementation.

\section*{Applicability of DeePC for nonlinear systems}

We will use the conventional DeePC described in (\ref{deepc}) as the foundation of developing our deep learning-enabled DeePC method. Given that (6) was originally developed for LTI systems based on Lemma~\ref{lemma1}, in this section, we leverage the concept of Koopman modeling for control to justify the suitability of using conventional DeePC in (\ref{deepc}) as the basis for developing DeePC-based controllers for nonlinear systems in (\ref{eq:nonlinear sys}).

Consider a discrete-time nonlinear control system described in (\ref{eq:nonlinear sys}). Based on the Koopman theory for controlled systems~\cite{korda2018linear}, the state vector can be extended to include control inputs, that is, $\chi_k = [x_k^\top, u_k^\top]^\top$. Accordingly, there exists an infinite-dimensional nonlinear lifting mapping $\Psi_\chi$ that allows the dynamics of the lifted state to be governed by a Koopman operator $\mathcal{K}$ as follows~\cite{proctor2018generalizing, korda2018linear, zhang2023reduced}:
\begin{equation}\label{paper1:koopman:linear:1}
	 \Psi_{\chi} ( \chi_{k+1}) = \mathcal{K} \Psi_{\chi} (\chi_k)
\end{equation}
where $\Psi_{\chi}(\chi_k) := [\Psi(x_k)^{\top}, u_k^{\top}]^{\top}$ with $\Psi$ being a nonlinear mapping for the system state $x$. 
The corresponding Koopman operator $\mathcal{K}$ can then be identified and represented as a block matrix as follows:\vspace{2mm}
\begin{equation}\label{K:block}
{\mathcal K} = \left[
    \begin{array}{c;{2pt/2pt}c}
        A_{\mathcal{K}}&B_{\mathcal{K}} \\ \hdashline[2pt/2pt]
        *&*
    \end{array}
\right]
\end{equation}\normalsize \vspace{1mm}

Since the primary focus is on the future behaviors of the lifted state $\Psi$ (which are directly related to the original system state $x$) instead of all the elements of $\Psi_\chi$, it is sufficient only to identify matrices $A_{\mathcal{K}}$ and $B_{\mathcal{K}}$. In addition, a projection matrix $D_{\mathcal{K}}$ is used to map the lifted state $\Psi$ back to the original state space. The Koopman-based linear model for nonlinear control system (\ref{eq:nonlinear sys}) can be described as follows~\cite{korda2018linear, zhang2023reduced, narasingam2020application, li2024data}:
\begin{subequations}\label{paper1:koopman:linear:2}
\begin{align}
    \Psi (x_{k+1}) &= A_{\mathcal{K}} \Psi (x_k) +B_{\mathcal{K}} u_k \\
    \hat{x}_k &= D_{\mathcal{K}}\Psi(x_k)
\end{align}
\end{subequations}

Let $z = \Psi(x) \in \mathbb{R}^{n_z}$ represent the lifted state vector. A Koopman-based LTI model can be constructed in the following form~\cite{cherd_koopman}: 
\begin{subequations}\label{eq:koopman deepc}
\begin{align}
     z_{k+1} &= A_{\mathcal{K}} z_k +B_{\mathcal{K}} u_k \\
    \hat{x}_k &= D_{\mathcal{K}}z_k \\
    y_k &=C_z z_k
\end{align}
\end{subequations}
where $C_z = C D_{\mathcal{K}}$ is the output matrix. (\ref{eq:koopman deepc}) characterizes (or more precisely, approximates) the dynamic behaviors of the original nonlinear system in~(\ref{eq:nonlinear sys}) in a higher-dimensional space. 

It is worth noting that the established Koopman-based LTI model in~(\ref{eq:koopman deepc}) has the same control input $u$ and controlled output $y$ as the underlying nonlinear system in~(\ref{eq:nonlinear sys}). This implies that for the nonlinear system in~(\ref{eq:nonlinear sys}), there exists a higher-dimensional LTI model that can represent or approximate the nonlinear dynamics of the underlying system, with control inputs and measured outputs remaining unaffected by coordinate changes, thereby retaining their original physical meaning. Meanwhile, the conventional DeePC method, originally developed for LTI systems in~(\ref{deepc}), may be directly applied to systems described by the Koopman-based model in~(\ref{eq:koopman deepc}) based on control input $u$ and controlled output $y$. Given that it is feasible to construct an accurate Koopman model in~(\ref{eq:koopman deepc}) as a high-fidelity surrogate for the nonlinear system in~(\ref{eq:nonlinear sys}), the DeePC scheme developed based on (\ref{eq:koopman deepc}) can serve as an effective predictive control solution for the nonlinear system in (\ref{eq:nonlinear sys}). The above analysis supports the suitability of applying DeePC for control of the nonlinear system in (\ref{eq:nonlinear sys}).



\section*{Deep learning-based DeePC control approach}

In this section, we propose the deep learning-based DeePC method -- Deep DeePC, which leverages deep learning to generate optimal control actions without the need to solve the optimization problem (\ref{deepc}) at each sampling instant during online implementation. In addition, the process of constructing a training dataset for Deep DeePC using historical offline data is explained.

\begin{figure*}[t!]
    \centering
    \includegraphics[width=1\textwidth]{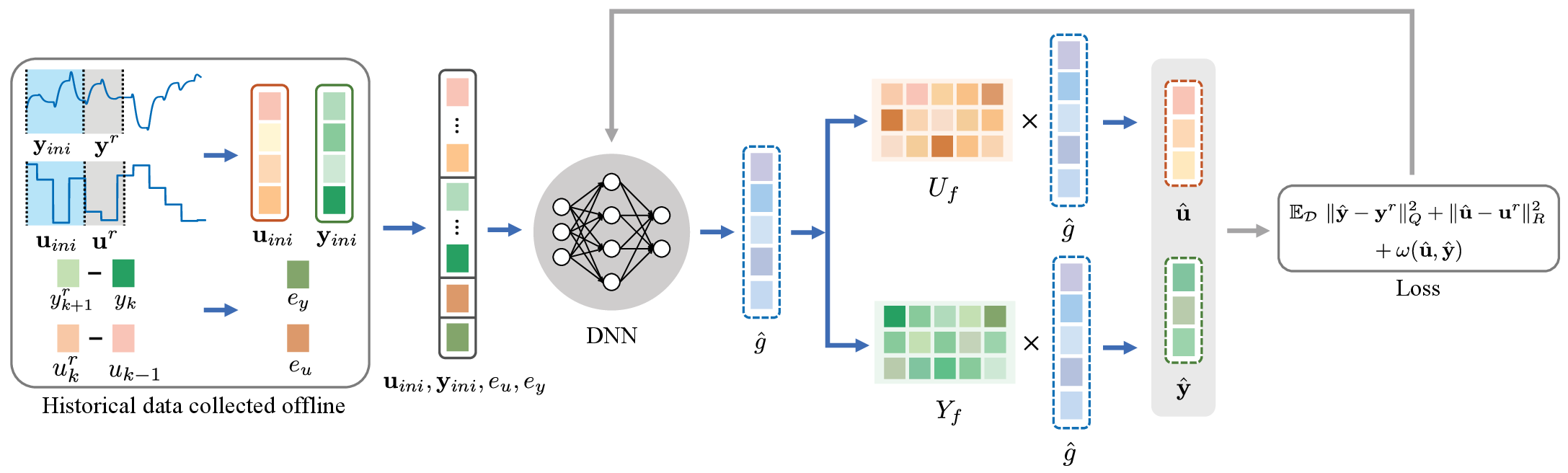}
    \caption{A graphical illustration of the proposed deep learning-enabled DeePC pipeline.}\label{ddeepc:architecture}
\end{figure*}

\subsection*{Structure of the proposed method}

A graphical illustration of the proposed Deep DeePC pipeline is presented in Figure \ref{ddeepc:architecture}. Historical open-loop data are utilized to construct a dataset, which is used to train a dense neural network (DNN). This DNN is used to output an approximation of the DeePC operator $g$ in (\ref{deepc}). The trajectories of future input and system output, denoted by $\hat{\mathbf{u}}$ and $\hat{\mathbf{y}}$, will be predicted based on the output of the DNN and (\ref{wfl-pf}). The objective is to train the DNN such that future system outputs are driven towards the reference trajectories.

\subsection*{Deep DeePC}\label{deepc control}

By incorporating the equality constraints $\hat{\mathbf{u}}_k=U_f g_k$ and $\hat{\mathbf{y}}_k=Y_f g_k$ from (\ref{deepc:2}) into the objective function (\ref{deepc:1}), the online optimization problem for DeePC in (\ref{deepc}) is re-formulated as follows~\cite{vahidi2023data}:
\begin{subequations}\label{deepc_2}
    \begin{align}
        \label{deepc_2:1} \min_{g_k} \  &\Vert Y_f g_k - \mathbf{y}^r_k \Vert_Q^2 + \Vert U_f g_k -\mathbf{u}^r_k \Vert_R^2 \\
        \text{s.t.} \quad &\left[\begin{array}{c}
            U_p \\
            Y_p 
          \end{array}
    \right] g_k = 
    \left[\begin{array}{c}
         \mathbf{u}_{ini,k} \\
         \mathbf{y}_{ini,k} 
    \end{array} \right]  \\
    &\hat{u}_{j|k} \in \mathbb{U}, \quad j = k, \ldots, k+N_p-1 \\
    &\hat{y}_{j|k} \in \mathbb{Y}, \quad j = k, \ldots, k+N_p-1
    \end{align}   
\end{subequations}
This way, the decision variable of the online optimization problem for DeePC has reduced from $g_k, \hat{\mathbf{u}}_k$, and $\hat{\mathbf{y}}_k$ to just the DeePC operator $g_k$.

In this work, we aim to realize online DeePC-based control of the nonlinear process in a manner that bypasses solving the online optimization in (\ref{deepc_2}). In the optimization problem (\ref{deepc_2}), the Hankel matrices $U_p$ and $Y_p$, constructed from offline open-loop process operation data, remain unchanged during each time instant update. Therefore, the elements of DeePC operator $g_k$, which are the decision variables of the DeePC formulation in (\ref{deepc_2}), are determined at each sampling instant $k, k \in \mathbb{Z}_{\geq 0}$, based on the initial trajectories of the inputs and the outputs, denoted by $\mathbf{u}_{ini,k}$ and $\mathbf{y}_{ini,k}$. 

As depicted in Figure~\ref{ddeepc:architecture}, a neural network, denoted by $F_{\theta}$, is trained, and the trained neural network will update the DeePC operator $g_k$ at each new sampling instant during online implementation. The structure of this neural network is designed as follows:
\begin{equation}\label{nn}
    \hat{g}_k = F_{\theta}(\mathbf{u}_{ini,k}, \mathbf{y}_{ini,k}, e_{u,k}, e_{y,k}|\theta)
\end{equation}
where $\hat{g}_k$ is the predicted DeePC operator, $\theta$ includes the trainable parameters of neural network $F_{\theta}$, $e_{u,k}=u^r_k - u_{k-1}$ is the input tracking error that represents the difference between the reference input at current instant $k$ and the most recent control input at time instant $k-1$, and $e_{y,k}=y^r_{k+1} - y_k$ is the output tracking error that represents the difference between the reference output at time instant $k+1$ and the latest measured output at time instant $k$. The time instants for the input and output tracking errors are different because the system output is one step ahead of the input. The inputs to $F_{\theta}$ include the initial input and output trajectories, that is, $\mathbf{u}_{ini,k}$ and $\mathbf{y}_{ini,k}$, which contain information for the past $T_{ini}$ steps. Since the neural network is used to generate future control inputs for set-point tracking, $e_{u,k}$ and $e_{y,k}$, which contain information for the most recent input and output values and their corresponding future references for the next step, are also incorporated as inputs to this neural network.
The dimension of the inputs to the neural network $F_{\theta}$ is determined based on the selected parameters $T_{ini}$ for DeePC design and the system parameters $n_u$ and $n_y$, resulting in a dimension of $(n_u + n_y) \times(T_{ini}+1)$. The dimension of the output of $F_{\theta}$ is the same as the dimension of DeePC operator $g_k$ in (\ref{deepc_2}), which is $T-T_{ini}-N_p+1$.

The objective of this neural network is to output DeePC operator $\hat{g}_k$, which can be further used to generate control actions that drive the system towards desired reference outputs -- this control objective aligns with that of DeePC (\ref{deepc_2:1}). 
To address the input and output constraints, a soft constraint $\omega(\hat{\mathbf{u}}, \hat{\mathbf{y}})$ is incorporated into the objective function. Given a dataset $\mathcal{D}$ composed of multiple $T_{ini}$-step initial input and output trajectories, $\mathbf{u}_{ini}$ and $\mathbf{y}_{ini}$, and their corresponding $N_p$-step reference input and reference output trajectories, $\mathbf{u}^r$ and $\mathbf{y}^r$, the objective function used to train the parameters in $\theta$ is as follows:
\begin{subequations}\label{loss}
    \begin{align}
        \mathcal{L} &= \mathbb{E}_{\mathcal{D}} \ \Vert \hat{\mathbf{y}} - \mathbf{y}^r \Vert_Q^2 + \Vert \hat{\mathbf{u}} - \mathbf{u}^r \Vert_R^2 + \omega(\hat{\mathbf{u}}, \hat{\mathbf{y}}) \\ 
         \label{loss:1-2} &=\mathbb{E}_{\mathcal{D}} \ \Vert Y_f \hat{g}    - \mathbf{y}^r \Vert_Q^2 + \Vert  U_f \hat{g}  -\mathbf{u}^r \Vert_R^2 + \omega(U_f \hat{g}, Y_f \hat{g})  
        \end{align}
\end{subequations}
where $\hat{\mathbf{u}}$ and $\hat{\mathbf{y}}$ are the predicted future input/output trajectories based on the DeePC operator $\hat{g}$ generated by the neural network. 
The soft constraint $\omega(\hat{\mathbf{u}}, \hat{\mathbf{y}})$ is described as follows:
\begin{equation}\label{w term}
        \omega(\hat{\mathbf{u}}, \hat{\mathbf{y}}) = \Vert \hat{\mathbf{u}} - \mathbf{u}_{lb} \Vert_{P'_{u, lb}}^2 +  \Vert  \mathbf{u}_{ub} - \hat{\mathbf{u}} \Vert_{P'_{u, ub}}^2 + \Vert \hat{\mathbf{y}} - \mathbf{y}_{lb} \Vert_{P'_{y, lb}}^2 +  \Vert  \mathbf{y}_{ub} - \hat{\mathbf{y}} \Vert_{P'_{y, ub}}^2
\end{equation}
where $\mathbf{u}_{lb} \in \mathbb{R}^{n_u N_p}$ and $\mathbf{u}_{ub} \in \mathbb{R}^{n_u N_p}$ are the lower and upper bounds of the input variables for $N_p$ steps, respectively (similarly for $\mathbf{y}_{lb}$ and $\mathbf{y}_{ub}$).  $P'_{u, lb} = P_u \odot M_{u, lb} \in \mathbb{R}^{n_u N_p \times n_u N_p }$ and $P'_{u, ub} = P_u \odot M_{u,ub} \in \mathbb{R}^{n_u N_p \times n_u N_p}$, where $P_u$ is the weighting matrix of the term that penalizes the violation of input constraint; $M_{u, lb} = \text{diag}([\mathbf{1}_{\hat{\mathbf{u}}(1) < \mathbf{u}_{lb}(1)}, \ldots, \mathbf{1}_{\hat{\mathbf{u}}(n_u N_p) < \mathbf{u}_{lb}(n_u N_p)}])$ and $M_{u, ub} = \text{diag}([\mathbf{1}_{\mathbf{u}_{ub}(1) < \hat{\mathbf{u}}(1)}, \ldots,$ $\mathbf{1}_{\mathbf{u}_{ub}(n_u N_p) < \hat{\mathbf{u}}(n_u N_p)}])$ are the mask matrices used to penalize the violated terms (similarly for $P'_{y,lb}$ and $P'_{y,ub}$).
 
The optimization problem associated with the offline neural network training can be formulated as follows:
\begin{equation}\label{ddeepc:opt}
    \min_{\theta} \mathcal L =\min_{\theta} \  \mathbb{E}_{\mathcal{D}} \  \Vert Y_f \hat{g}    - \mathbf{y}^r \Vert_Q^2 + \Vert  U_f \hat{g} -\mathbf{u}^r \Vert_R^2 + \omega(U_f \hat{g}, Y_f \hat{g})  
\end{equation}
where $\hat{g} = F_{\theta}(\mathbf{u}_{ini}, \mathbf{y}_{ini}, e_u, e_y |\theta)$.

Once the training is completed, the trained neural network $F_{\theta}$ is used to output DeePC operator $\hat{g}_k$ online following (\ref{nn}), at each sampling time instant $k$, $k\in\mathbb{Z}_{\geq 0}$. The DeePC operator $\hat{g}_k$ is then used to generate the optimal control sequence $\hat{\mathbf{u}}^*_k$ following: 
\begin{equation}\label{eq:deep deepc 1}
    \hat{\mathbf{u}}^*_k = U_f \hat{g}_k = U_f  F_{\theta}(\mathbf{u}_{ini, k}, \mathbf{y}_{ini, k}, e_{u, k}, e_{y, k}| \theta^*)
\end{equation}
where $\theta^*$ is the optimal neural network parameters after training. The first control action $\hat{u}_{k|k}^*$ in $\hat{\mathbf{u}}^*_k$ will be applied to the system to achieve desired control objectives. It is worth noting that the proposed Deep DeePC method is free of online optimization, once the neural network is fully trained offline.

\begin{rmk}
    In the existing literature, neural networks have been utilized to approximate the control policy of MPC controllers~\cite{cao2020deep, li2022using}, which is another effective framework to substantially increase the online computation speed of optimal control. This type of approach typically relies on an existing MPC controller developed based on a dynamic model to generate closed-loop data for neural network training. However, when a dynamic model is unavailable and only input and output data are accessible, the proposed Deep DeePC method can serve as a viable alternative.
\end{rmk}


\subsection*{Training data construction}

\begin{figure*}[t!]
    \centering
    \includegraphics[width=1\textwidth]{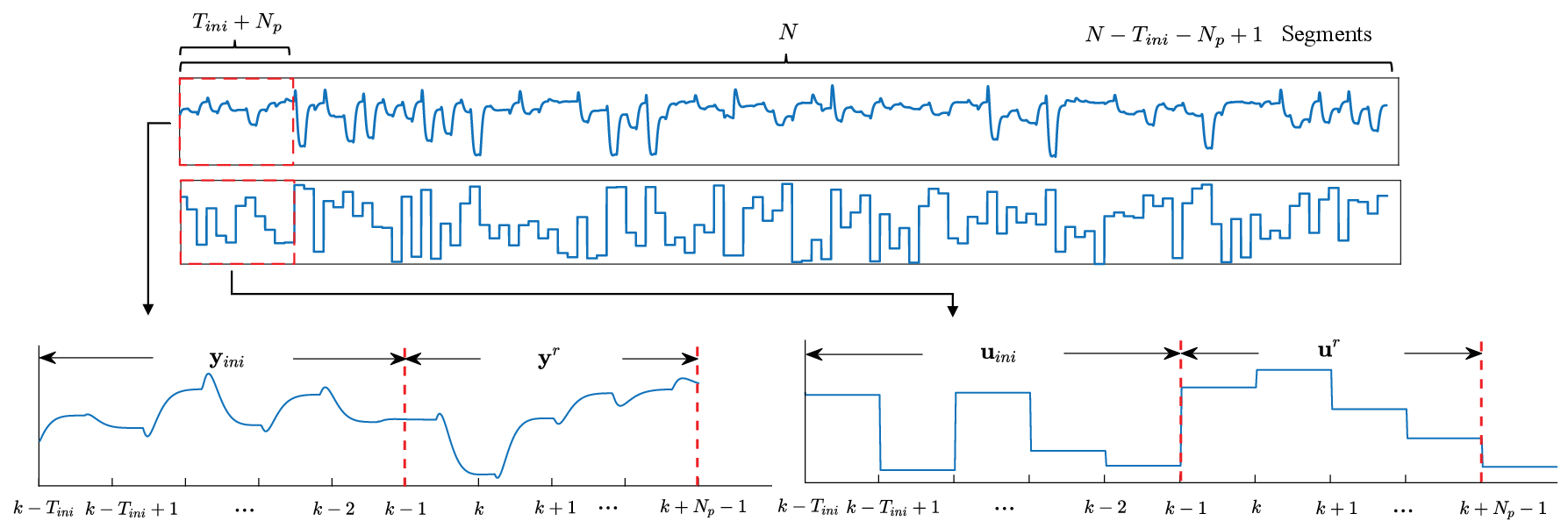}
    \caption{A graphical illustration of the construction of training data based on system historical data collected offline.}\label{fig:data}
\end{figure*}

Training the neural network in (\ref{nn}) based on the objective function in (\ref{loss}) requires a dataset that comprises initial input and initial output trajectories (denoted by $\mathbf{u}_{ini}$ and $\mathbf{y}_{ini}$, respectively), and their corresponding future reference input and reference output trajectories (denoted by $\mathbf{u}^r$ and $\mathbf{y}^r$, respectively). The initial input and output trajectories are used as inputs to the neural network in (\ref{nn}), while the future references serve as the labels that are used to compute the values of the objective function, as needed for supervised learning-based training as described in (\ref{loss}). 
The combined initial trajectories and future references should consist of samples collected at consecutive time instants that span a specific $(T_{ini} + N_p)$-step time window. In this work, we utilize historical data to construct the input and label data for training the neural network.

Figure \ref{fig:data} provides a graphical illustration of the construction of a training dataset based on historical data of the system collected offline. The objective is to construct $(T_{ini}+N_p)$-step continuous sequences, which can then be divided into input and label data for neural network training. As shown in Figure~\ref{fig:data}, $(N - T_{ini} - N_p + 1)$ segments of the  $(T_{ini}+N_p)$-step sequences can be extracted from an $N$-step sequential system trajectory. Within each segment, the first $T_{ini}$ steps of the segment are treated as the initial trajectories $\mathbf{u}_{ini}$ and $\mathbf{y}_{ini}$, which are the inputs of the neural network, and the remaining $N_p$ steps of the segment are regarded as the future references $\mathbf{u}^r$ and $\mathbf{y}^r$ corresponding to the initial trajectories, which are the labels of the supervised learning. In addition, the input and output errors, $e_u$ and $e_y$, are computed by the constructed initial trajectories and future references of the selected segment based on (\ref{nn}).


\section*{Event-based constraint handling}

\begin{figure*}[t!]
    \centering
    \includegraphics[width=1\textwidth]{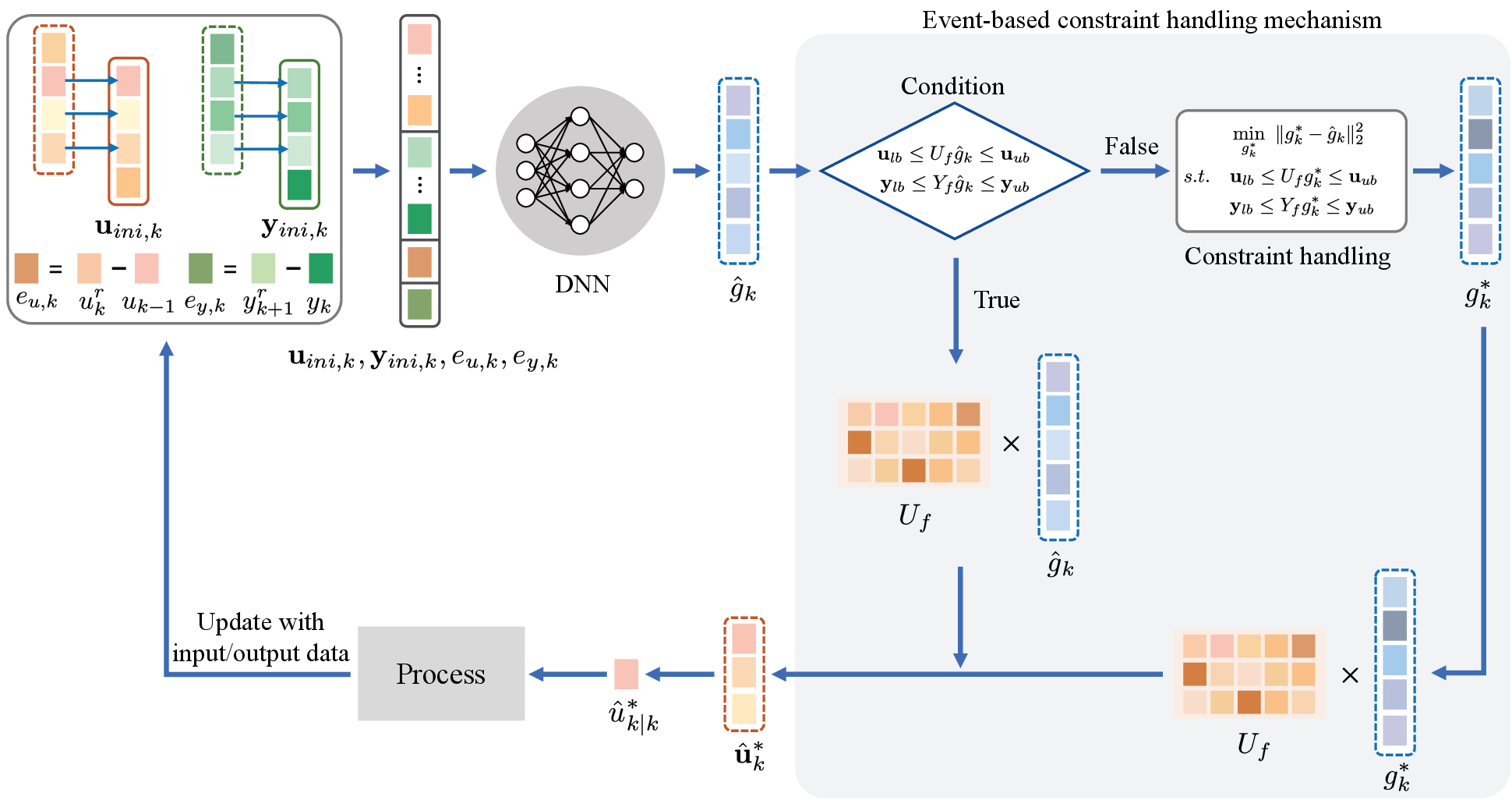}
    \caption{A block diagram of the online implementation of the proposed deep learning-enabled DeePC design with an event-based constraint handling scheme.}\label{fig:online}
\end{figure*}

The proposed Deep DeePC approach in the ``Deep DeePC'' section is highly computationally efficient since it does not require solving online optimization. Meanwhile, it is worth mentioning that the Deep DeePC controller in (\ref{eq:deep deepc 1}) is not capable of handling hard constraints on either the system inputs or the system outputs. To deal with cases when constraint satisfaction is critical, for example, when safety-related constraints need to be satisfied, we further propose an event-based constraint handling scheme. This scheme can be incorporated into the proposed online optimization-free Deep DeePC (\ref{eq:deep deepc 1}) to ensure the satisfaction of input and output constraints during online implementation, when it is necessary. An illustrative diagram of the Deep DeePC approach with the constraint handling scheme is presented in Figure \ref{fig:online}. 

The online implementation of Deep DeePC with the constraint handling scheme is described in Algorithm \ref{alg}. At each sampling instant $k$, $k\in\mathbb{Z}_{\geq 0}$, the trained DNN model generates the DeePC operator $\hat{g}_k$ based on the $\mathbf{u}_{ini,k}$, $\mathbf{y}_{ini,k}$, $e_{u,k}$, and $e_{y,k}$ collected during online operation:
\begin{equation}\label{deepc:neural:3}
    \hat{g}_k= F_{\theta}(\mathbf{u}_{ini,k}, \mathbf{y}_{ini,k}, e_{u,k}, e_{y,k}| \theta^*)
\end{equation}
Then future optimal control sequence  $\hat{\mathbf{u}}_k$ can be obtained based on (\ref{eq:deep deepc 1}), and the corresponding future output trajectories $\hat{\mathbf{y}}_k$ can be computed following DeePC framework, i.e., $\hat{\mathbf{y}}_k = Y_f \hat{g}_k$.

\begin{algorithm}[t!]
\vspace{1.5mm}
\caption{Online implementation of Deep DeePC with constraint handling scheme}\label{alg}
\KwIn{Trained neural network $F_{\theta}$ with optimized parameters $\theta^*$; established Hankel matrices $U_f$ and $Y_f$ based on offline collected data $\mathbf{u}^d_T$ and $\mathbf{y}^d_T$; steady-state input and reference outputs $\mathbf{u}^r$ and $\mathbf{y}^r$.}  \vspace{2.5mm}

\KwOut{Real-time input and output trajectories.} \vspace{2.5mm}

\SetAlgoLined

{Initialize $\mathbf{u}_{ini,T_{ini}}$, $\mathbf{y}_{ini, T_{ini}}$, $e_{u, T_{ini}}$, and $e_{y, T_{ini}}$.} \vspace{2.5mm}

\parbox[t]{0.92\linewidth}{\While{$k > T_{ini}$}{
    \vspace{2.5mm}
    \begin{enumerate}[label=\textbf{\scriptsize 2.\arabic*}]
        \item Compute the DeePC operator following $\hat{g}_k= F_{\theta}(\mathbf{u}_{ini,k}, \mathbf{y}_{ini,k}, e_{u,k}, e_{y,k}| \theta^*)$.
        \item \parbox[t]{0.9\linewidth}{\eIf{$\mathbf{u}_{lb} \leq U_f \hat{g}_k \leq \mathbf{u}_{ub} \ \text{and} \ \mathbf{y}_{lb} \leq Y_f \hat{g}_k \leq \mathbf{y}_{ub}$}{ \vspace{2.5mm}
            Compute the optimal control sequence following $\hat{\mathbf{u}}_k^* = U_f \hat{g}_k$. \vspace{1.5mm}
        }{ \vspace{2.5mm}
            {Solve optimization problem (\ref{opt:g}) to update DeePC operator $g^*_k$.} \vspace{2.5mm}
            
            {Compute the optimal control sequence following $\hat{\mathbf{u}}_k^* = U_f g^*_k$.} \vspace{2.5mm}
        }}
        
        \item Apply $u_k$ where $u_k = \hat{u}^*_{k|k}$ to the process in (\ref{eq:nonlinear sys}). 

        \item Update $\mathbf{u}_{ini, k+1} := \{u\}^{k}_{k-T_{ini}+1}$, $\mathbf{y}_{ini, k+1} := \{y\}^{k}_{k-T_{ini}+1}$.

        \item Update $e_{u, k+1} = u^r_{k+1} - u_{k}$, $e_{y, k+1} = y^r_{k+2} - y_{k+1}$.

        \item $k \leftarrow k+1$.
    
    \end{enumerate}
}}
\end{algorithm}

The predicted future input and output trajectories generated based on $\hat g_{k}$ in (\ref{deepc:neural:3}) may not always satisfy the hard constraints. In such cases, if any of the predicted future states violate the constraints, a constraint handling scheme is proposed to adjust the DeePC operator $\hat{g}_k$ to comply with the constraints. The objective of this design is to find an optimized DeePC operator $g^*_k$ which is close to $\hat{g}_k$ while ensuring that the future states meet the system constraints. The optimization problem is formulated as follows:
\begin{subequations}\label{opt:g}
    \begin{align}
        &\min_{g^*_k} \ \Vert g^*_k - \hat{g}_k\Vert_2^2\\
        \text{s.t.} \quad & \mathbf{u}_{lb} \leq U_f g^*_k \leq \mathbf{u}_{ub} \\
        & \mathbf{y}_{lb} \leq Y_f g^*_k \leq \mathbf{y}_{ub}
    \end{align}
\end{subequations}

We note that if the predicted future $N_p$-step input and output trajectories do not violate the system constraints, then (\ref{opt:g}) does not need to be solved, and the optimal control sequence generated based on (\ref{eq:deep deepc 1}) can be directly obtained based on $\hat{g}_k$. The optimal control sequence $\hat{\mathbf{u}}^*_k$ for $N_p$ steps can be obtained and described as follows:
\begin{equation}\label{eq:cdeep deepc 1}
    \hat{\mathbf{u}}^*_k = 
    \begin{cases}
         U_f \hat{g}_k, & \text{if} \ \mathbf{u}_{lb} \leq U_f \hat{g}_k \leq \mathbf{u}_{ub} \ \text{and} \ \mathbf{y}_{lb} \leq Y_f \hat{g}_k \leq \mathbf{y}_{ub}\\
         U_f g^*_k,  &\text{otherwise}
    \end{cases}
\end{equation}
The first control action $\hat{u}^*_{k|k}$ in the optimal control sequence $\hat{\mathbf{u}}^*_k$ is applied to the system to achieve the desired control performance.

\section*{Extension of the proposed method}
In this section, we present an alternative method to extend our proposed Deep DeePC approach to handle situations where steady-state reference inputs are unavailable for the nonlinear process under consideration.


The online implementation of the well-trained Deep DeePC model for control tasks requires information on the initial input and output trajectories, $\mathbf{u}_{ini}$ and $\mathbf{y}_{ini}$, along with the input and output errors, $e_u$ and $e_y$. However, in some cases where the steady-state reference inputs corresponding to the set-points are unknown, implementing the current approach becomes impractical. Therefore, an alternative design for Deep DeePC that does not require steady-state reference inputs is proposed.

The neural network is described as follows:
\begin{equation}\label{nn:2}
    \hat{g}_k = \tilde{F}_{\tilde{\theta}}(\mathbf{u}_{ini,k}, \mathbf{y}_{ini,k}, e_{y,k}|\tilde{\theta})
\end{equation}
where $\tilde{F}_{\tilde{\theta}}$ denotes the neural network without requiring the steady-state reference input; $\tilde{\theta}$ denotes the trainable parameters of neural network $\tilde{F}_{\tilde{\theta}}$. The input dimension of the neural network $\tilde{F}_{\tilde{\theta}}$ is reduced to $(n_u + n_y) \times T_{ini} + n_y$. Given a dataset $\tilde{\mathcal{D}}$ that is composed of multiple $\mathbf{u}_{ini}$, $\mathbf{y}_{ini}$, and corresponding $\mathbf{y}^r$,  the objective function used to train the neural network $\tilde{F}_{\tilde{\theta}}$ is defined as follows:
\begin{subequations}\label{loss:2}
    \begin{align}
        \tilde{\mathcal{L}} &= \mathbb{E}_{\tilde{\mathcal{D}}} \ \Vert \hat{\mathbf{y}} - \mathbf{y}^r \Vert_Q^2 + \omega(\hat{\mathbf{u}}, \hat{\mathbf{y}})  \\
        &= \mathbb{E}_{\tilde{\mathcal{D}}} \ \Vert Y_f \hat{g} - \mathbf{y}^r \Vert_Q^2 + \omega(U_f \hat{g}, Y_f \hat{g})
    \end{align}
\end{subequations}

The optimization problem for training of neural network $\tilde{F}_{\tilde{\theta}}$ without the steady-state reference inputs can be formulated as follows:
\begin{equation}\label{ddeepc:opt:2}
     \min_{\tilde{\theta}} {\tilde {\mathcal L }}= \min_{\tilde{\theta}} \ \mathbb{E}_{\tilde{\mathcal{D}}} \  \Vert Y_f \hat{g} - \mathbf{y}^r \Vert_Q^2 + \omega(U_f \hat{g}, Y_f \hat{g})
\end{equation}
where $\hat{g} = \tilde{F}_{\tilde{\theta}}(\mathbf{u}_{ini}, \mathbf{y}_{ini}, e_y|\tilde{\theta})$.

The trained neural network $\tilde{F}_{\tilde{\theta}}$ is used to output DeePC operator $\hat{g}_k$, at each sampling time instant $k$, $k\in\mathbb{Z}_{\geq 0}$. The DeePC operator is used to generate the optimal control sequence $\hat{\mathbf{u}}_k^*$ following:
\begin{equation}\label{eq:deep deepc 2}
    \hat{\mathbf{u}}_k^* = U_f \hat{g}_k = U_f \tilde{F}_{\tilde{\theta}} (\mathbf{u}_{ini, k}, \mathbf{y}_{ini, k}, e_{y, k}| \tilde{\theta}^*)
\end{equation}
where $\tilde{\theta}^*$ is the well-trained parameters of neural network $\tilde{F}_{\tilde{\theta}}$. The first control action $\hat{u}_{k|k}^*$ in $\hat{\mathbf{u}}_k^*$ will be applied to drive the system to the desired operation condition.

\section*{Case study on gene regulatory network}

\subsection*{Process description}

A gene regulatory network (GRN) is a nanoscale dynamic system within synthetic biology~\cite{elowitz2000synthetic, han2021desko}. We consider a three-gene regulatory network, where both mRNA and protein dynamics display oscillatory behaviors. The transcription and translation dynamics within this GRN can be described by a discrete-time nonlinear system, given as follows~\cite{elowitz2000synthetic, han2021desko}:
\begin{subequations}\label{eq:grn}\small
    \begin{align}
        x_{k+1}(i) &= x_k(i) + \left ( - \gamma_i x_k(i) + \frac{a_i}{K_i + x_k^2(j)} + u_k(i) \right) \cdot \Delta + \xi_k (i), \ (i, j) \in \{(1,6); (2,4); (3,5)\} \\
        x_{k+1}(i) &= x_k(i) + \left( - c_j x_k (i) + \beta_j x_k (j) \right) \cdot \Delta  + \xi_k (i), \ (i, j) \in \{(4,1); (5,2); (6,3)\}
    \end{align}
\end{subequations}
where $x(i), i=1,2,3$, denote the concentrations of the mRNA transcripts for the three different genes; $x(i), i=4,5,6$, denote the concentrations of the corresponding proteins for three genes; $\xi(i), i=1, \ldots, 6$, denote the independent and identically distributed (i.i.d.) uniform noise, that is, $\xi(i) \thicksim \mathcal{U}(-\delta, \delta)$; $u(i), i=1,2,3$, denote the number of protein copies per cell produced from a given promoter type during continuous growth; $a_i, i=1,2,3$, denote the maximum promoter strength for their corresponding gene; $\gamma_i, i=1,2,3$, denote the mRNA degradation rates; $c_i, i=1,2,3$, denote the protein degradation rates; $\beta_i, i=1,2,3$, denote the protein production rates; $K_i, i=1,2,3$, denote the dissociation constants; $\Delta$ denotes the discretization time step, which is also the sampling period.

The state variables of the system include the concentrations of the mRNA ($x(i), i=1,2,3$) and the concentrations of the proteins ($x(i), i=4,5,6$). The measured output variables are the protein concentrations $x(i), i=4,5,6$, which are measured online using fluorescent markers. The control inputs $u(i), i=1,2,3$, are light control signals that can induce the expression of genes through the activation of their photo-sensitive promoters.

\subsection*{Settings}

\subsubsection*{Control methods being evaluated}

In this section, three controllers are developed based on the following three control approaches: the proposed Deep DeePC method, which is referred to as Deep DeePC as described by (\ref{eq:deep deepc 1}); the proposed Deep DeePC method with the proposed constraint handling scheme, which is referred to as constrained Deep DeePC as described by (\ref{eq:cdeep deepc 1}); and conventional DeePC, which is referred to as DeePC as described in (\ref{deepc}).

\subsubsection*{Parameters}

The following parameters are chosen for the three controllers to ensure a fair comparison: $T=200$, $T_{ini}=10$, $N_p=10$, $Q=5 \times I_{30}$, and $R=1 \times I_{30}$. Deep DeePC and constrained Deep DeePC use the same weighting matrices to penalize the violation of input and output constraints; specifically, $P_u = 10 \times I_{30}$ and $P_y = 10 \times I_{30}$, respectively. The Hankel matrices $U_p$, $U_f$, $Y_p$, and $Y_f$ are identical across all designs.

The neural network architecture accounting for the DeePC operator consists of an input layer, two hidden layers, and an output layer. The number of neurons in different layers of the neural network involved in Deep DeePC is 66-150-150-181. The network uses the rectified linear unit (ReLU) as the activation function after the input and hidden layers. The training process involves 1000 epochs with a batch size of 200. The Adam optimizer, with a learning rate of $10^{-4}$, is employed for training. To ensure a fair comparison, the constrained Deep DeePC model utilizes the same well-trained neural network as the Deep DeePC.

\subsubsection*{Simulation setting}

In this study, the following system parameters are adopted for the GRN process: $K_i=1$, $a_i=1.6$, $\gamma_i=0.16$, $\beta_i=0.16$, $c_i=0.06$ $(i=1,2,3)$, $\delta=0$, and $\Delta=1$ min, which are the same as those in Reference~\cite{han2021desko}. First, open-loop simulations are conducted using the first-principles model (\ref{eq:grn}) to generate data. The generated data are used for two purposes: 1) to construct the Hankel matrices $U_p, U_f, Y_p,$ and $Y_f$, and 2) to train the DeePC operator within the proposed framework. $T=200$ is used as we construct the Hankel matrices. The data size for training the neural network is $10^4$. The control inputs $u(i), i=1,2,3$, are generated randomly following a uniform distribution with a prescribed range. Particularly, $u(i) \in [0,1], i=1,2,3$, and each of the inputs varies every 30 sampling periods.

During the online implementation, four open-loop stable steady states are considered as the set-points (i.e., the reference outputs) for the proposed controller to track. The tracking set-points are varied after every 100 sampling periods. The initial state $x_0$ and the four set-points $x_{si}, i=1,\ldots, 4$, are listed in Table \ref{table:grn:sp:x}. The control inputs $u_{si}, i=1, \ldots, 4$, corresponding to the four set-points are listed in Table \ref{table:grn:sp:u}.

\begin{table*}[tb!]
  \renewcommand\arraystretch{1.25}
  \caption{The initial state $x_0$ and the four set-points $x_{si}, i=1, \ldots, 4$, of GRN.}\label{table:grn:sp:x}\vspace{5pt}
  \centering
    \begin{tabular}{ c c c c c c c  }
      \toprule
       &  $x(1)$ &  $x(2)$ &  $x(3)$ &  $x(4)$ &  $x(5)$ &  $x(6)$  \\
      \midrule
      $x_{0}$ &
        29.36 & 24.60 & 20.30 &
        78.29 & 65.61 & 54.12 \\
      $x_{s1}$ &
        22.47 & 17.29 & 11.65 &
        59.93 & 46.10 & 31.06 \\
      $x_{s2}$ &
        15.85 & 14.44 & 15.60 &
        42.27 & 38.51 & 41.59 \\
      $x_{s3}$ &
        16.67 & 70.86 & 13.26 &
        44.45 & 18.90 & 35.35 \\  
      $x_{s4}$ &
        62.49 & 82.61 & 18.09 &
        16.66 & 22.03 & 48.24 \\
      \bottomrule
    \end{tabular}
\end{table*}

\begin{table}[t!]
  \renewcommand\arraystretch{1.25}
  \caption{Steady-state control inputs $u_{si}, i=1, \ldots, 4$, of GRN.}\label{table:grn:sp:u}\vspace{5pt}
  \centering
    \begin{tabular}{ c c c c c c c c c c }
      \toprule
      & $u(1)$ & $u(2)$ & $u(3)$  \\
      \midrule
      $u_{s1}$ &
      0.7189 & 0.5536 & 0.3725 \\
      $u_{s2}$ &
      0.5070 & 0.4620 & 0.4989 \\
      $u_{s3}$ &
      0.5332 & 0.2266 & 0.4233 \\
      $u_{s4}$ &
      0.1998 & 0.2632 & 0.5783 \\
      \bottomrule
    \end{tabular}
\end{table}

\subsection*{Control performance}

\begin{figure*}[t!]
    \centering
    \subfigure[Output trajectories based on Deep DeePC and constrained Deep DeePC.]{
    \label{fig:grn:y}
    \includegraphics[width=1\textwidth]{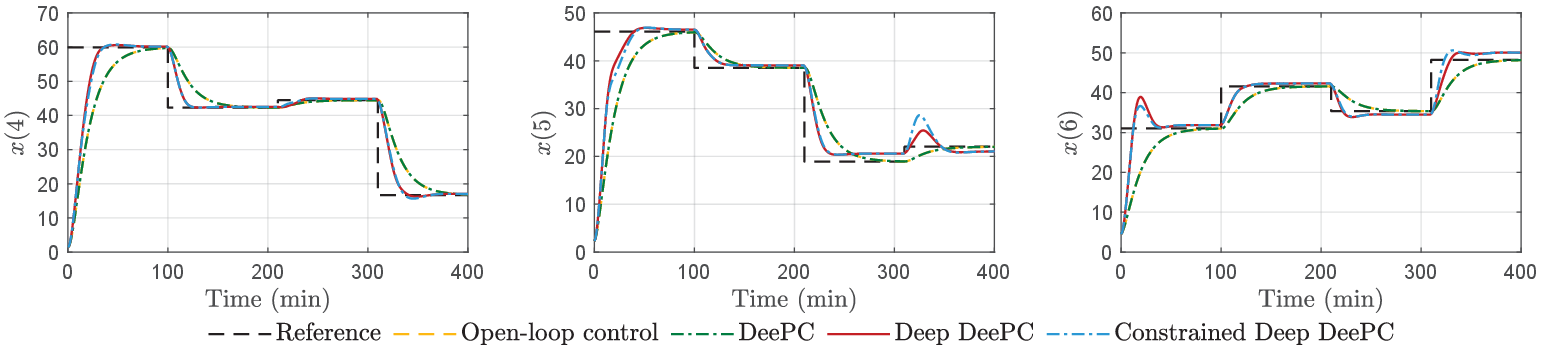}
    }

    \subfigure[Control input trajectories based on Deep DeePC and constrained Deep DeePC.]{
    \label{fig:grn:u}
    \includegraphics[width=1\textwidth]{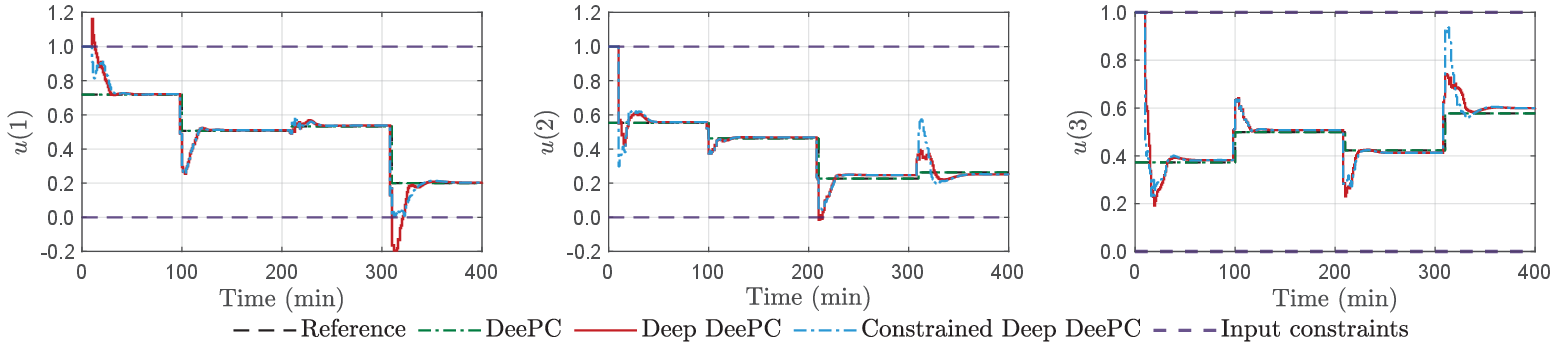}
    }
    \caption{Output and control input trajectories of GRN under designs with reference input.}
    \label{fig:grn:yu}
\end{figure*}

Figure \ref{fig:grn:yu} shows the closed-loop output trajectories and control input trajectories obtained using Deep DeePC, constrained Deep DeePC, and conventional DeePC. Figure \ref{fig:grn:y} illustrates the closed-loop output trajectories for the three controllers, alongside the open-loop output trajectories with steady-state reference inputs. Figure \ref{fig:grn:u} displays the control input trajectories generated by the different controllers. All three DeePC designs can drive the system to the desired set-points. The proposed Deep DeePC (shown in red lines) and constrained Deep DeePC (shown in blue dashed lines) achieve faster convergence compared to conventional DeePC. In addition, as shown in Figure \ref{fig:grn:u}, the proposed Deep DeePC occasionally violates the input constraints, while the proposed constraint handling scheme ensures that the system constraints are satisfied throughout the entire online implementation.

To quantitatively assess the control performance, we calculate the root mean squared error (RMSE) for each controller. The RMSE is defined as $\text{RMSE}=\sqrt{\frac{1}{n_y N_K}\sum_{k=1}^{N_K} \Vert y^r_k - y_k\Vert_2^2}$, where $N_K$ is the total number of sampling instants throughout the system operation. The RMSEs are computed using scaled system outputs. Table \ref{table:grn:rmse} presents the RMSE results for the three controllers, which demonstrate that both the proposed Deep DeePC method and the constrained Deep DeePC method outperform the conventional DeePC.

\begin{table}[t!]
  \renewcommand\arraystretch{1.25}
  \caption{Control performance comparison of GRN in terms of RMSEs.}\label{table:grn:rmse}\vspace{5pt}
  \centering\small
    \begin{tabular}{c c c c c }
      \toprule
      & Open-loop & DeePC & Deep DeePC & \thead{Constrained \\ Deep DeePC} \\ 
      \midrule
      RMSE & 0.17217 & 0.17218 & 0.14062 & 0.14182 \\
      \bottomrule
    \end{tabular}
\end{table}

\section*{Case study on chemical process}

In this section, we apply the proposed approach to a reactor-separator process. We examine two scenarios: when steady-state reference inputs are available and when they are unavailable.

\subsection*{Description of the process}

\begin{figure}[t]
    \centering
    \includegraphics[width=0.8\textwidth]{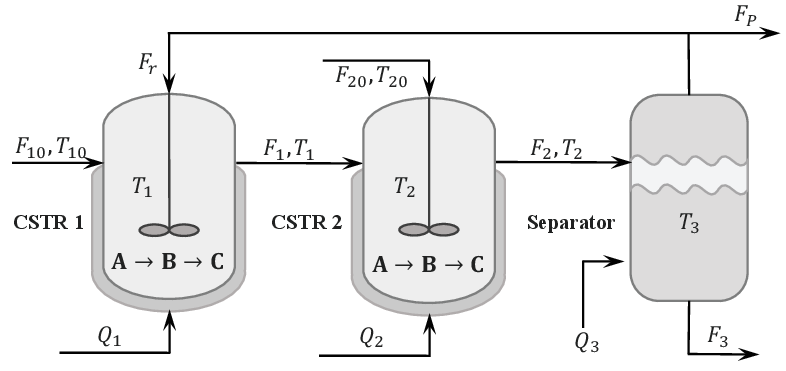}
    \caption{A schematic diagram of the reactor-separator process.}
    \label{3tanks}
\end{figure}

This reactor-separator process consists of two continuous stirred tank reactors (CSTRs) and one flash tank separator. A schematic diagram of this process is presented in Figure~\ref{3tanks}. This process involves two irreversible reactions: reactant $\mathbf{A}$ converts into the desired product $\mathbf{B}$, and $\mathbf{B}$ converts into the undesired side product $\mathbf{C}$, simultaneously.

In this process, the three vessels are interconnected through mass and energy flows. In the first reactor (CSTR 1), the input is pure reactant $\mathbf{A}$ at temperature $T_{10}$ and volumetric flow rate $F_{10}$. In the second reactor (CSTR 2), the input includes the output stream from CSTR 1 and another stream containing pure $\mathbf{A}$ at temperature $T_{20}$ and flow rate $F_{20}$. The effluent of CSTR 2 is sent to the separator with temperature $T_2$ and flow rate $F_2$. The state variables of the reactor separator process include the mass fractions of reactant $\mathbf{A}$ ($x_{\mathbf{A}i}, i=1, 2,3$), the mass fractions of the desired product $\mathbf{B}$ ($x_{\mathbf{B}i}, i=1, 2,3$), and the temperature ($T_{i}, i=1, 2,3$) in three vessels. The control inputs are the heat input rate $Q_i, i=1,2,3$, to the three vessels. Typically, only the temperature $T_i, i=1,2,3$, in three vessels are measured online using sensors. A detailed process description and the first-principles model of this process, which is used as the process simulator, can be found in References~\cite{zhang2023reduced, liu2009distributed}. The objective is to implement the proposed Deep DeePC approach to drive the process operation towards desired set-points by adjusting the heat input $Q_i, i=1,2,3$, to the three vessels.

\subsection*{Problem formulation and simulation settings}

\subsubsection*{Evaluated control methods}

We consider five control methods to illustrate the effectiveness and superiority of the proposed methods: the proposed Deep DeePC method that uses steady-state reference inputs for offline training and online control implementation, which is referred to as Deep DeePC-I as described by (\ref{eq:deep deepc 1}); the proposed Deep DeePC method that does not require steady-state reference inputs for its online implementation, which is referred to as Deep DeePC-II as described by (\ref{eq:deep deepc 2}); Deep DeePC-I with the proposed constraint handling scheme, which is referred to as constrained Deep DeePC-I as described by (\ref{eq:cdeep deepc 1}); Deep DeePC-II with the proposed constraint handling scheme, which is referred to as constrained Deep DeePC-II; and conventional DeePC, which is referred to as DeePC as described in (\ref{deepc}).

\subsubsection*{Parameter settings}

The following parameters are chosen for all the five control methods to ensure a fair comparison: $T = 200$, $T_{ini}=10$, $N_p=10$, $Q=5 \times I_{30}$, and $R=I_{30}$. The four proposed Deep DeePC designs employ the same weighting matrices to penalize the violation of input and output constraints, which are $P_u = 10 \times I_{30}$ and $P_y = 10 \times I_{30}$, respectively. The Hankel matrices $U_p, U_f, Y_p,$ and $Y_f$ remain the same for all the DeePC-based controllers.

The neural networks of Deep DeePC-I and Deep DeePC-II share a similar structure. Each neural network contains four layers: an input layer, two hidden layers, and an output layer. Both neural networks utilize the 
ReLU function as the activation function following the input and hidden layers. The training epoch is set to 1000, and the batch size is 200. The optimizer for the neural network training is Adam, and the learning rate is set as $10^{-4}$. The numbers of neurons in different layers of the neural networks involved in Deep DeePC-I and Deep DeePC-II are 66-150-150-181 and 63-150-150-181, respectively. For fair comparisons, constrained Deep DeePC-I uses the same well-trained neural network used in Deep DeePC-I (similarly to constrained Deep DeePC-II).

\begin{rmk}
    In this work, we primarily focus on integrating neural networks with conventional DeePC to achieve efficient online computation. In the two case studies, we employed fully-connected feedforward deep neural networks. Each of the neural networks in the two case studies has two hidden layers. When dealing with systems with larger scales, the number of hidden layers may need to be increased.
\end{rmk}

\begin{rmk}
    The control performance of the proposed Deep DeePC approach is dependent on several tuning parameters, including the length of historical data, the length of initial trajectories, the length of future prediction trajectories, and weighting matrices in~(\ref{loss:1-2}). These parameters need to be tuned through trial-and-error. If the control performance of Deep DeePC remains unsatisfactory, it is likely due to insufficient training data for the considered system. In such cases, acquiring additional data for neural network training may be necessary.
\end{rmk}

\subsubsection*{Data generation and process setting}

\begin{table}[t!]
  \renewcommand\arraystretch{1.25}
  \caption{The upper bounds and lower bounds of control inputs and controlled outputs of reactor-separator process.}\label{table:bound}\vspace{5pt}
  \centering
    \begin{tabular}{ c c c c c c c c c c c c c}
      \toprule
        & $Q_1~(\text{kJ/h})$ & $Q_2~(\text{kJ/h})$ & $Q_3~(\text{kJ/h})$ & $T_1$ (K) & $T_2$ (K) &$T_3$(K) \\
      \midrule
      Upper bounds &$3.1 \times 10^6$ & $1.3 \times 10^6$ & $3.1 \times 10^6$ &494.0 &486.0 &488.0 \\
      Lower bounds &$2.6 \times 10^6$ & $7.0 \times 10^5$ & $2.6 \times 10^6$ &480.0 &472.0 &474.0\\
      \bottomrule
    \end{tabular}
\end{table}

First, open-loop process simulations are conducted using the first-principles model of the reactor-separator process to generate data. The generated batch open-loop data are divided into two parts for different tasks: one part is used to establish Hankel matrices $U_p, U_f, Y_p,$ and $Y_f$ for the DeePC design; the rest of the data are used for training the proposed controllers. The data size used for constructing Hankel matrices is determined by the parameter $T=200$, and the size of the training data is $10^4$. The sampling period is $0.025$ hours. The heat inputs $Q_i, i=1,2,3$, are generated randomly in a uniform distribution with the prescribed ranges and are varied every 2 hours. The upper bounds and lower bounds of the heat inputs and temperature outputs are shown in Table \ref{table:bound}. Disturbances with bounded ranges are introduced into the process. The disturbances added to the mass fraction ($x_{\mathbf{A}i}$ and $x_{\mathbf{B}i}, i=1,2,3$) and temperature ($T_i, i=1,2,3$) follow Gaussian distribution with zero mean and standard deviation of 1 and 5, respectively. The bounds for the disturbances in mass fraction and temperature are set to $[-0.5, 0.5]$ and $[-5, 5]$, respectively.

During the online implementation, five open-loop stable steady states are considered as the set-points (i.e., the reference outputs) to illustrate our proposed methods. The tracking set-points are varied after every 2.5 hours. The initial state $x_0$ and the five set-points $x_{si}, i=1, \ldots, 5$, are presented in Table \ref{table:sp:x}. The control inputs $u_{si}, i=1, \ldots, 5$, corresponding to the five set-points are presented in Table \ref{table:sp:u}.

\begin{table*}[tb!]
  \renewcommand\arraystretch{1.25}
  \caption{The initial state $x_0$ and the five set-points $x_{si}, i=1, \ldots, 5$, of reactor-separator process.}\label{table:sp:x}\vspace{5pt}
  \centering
    \begin{tabular}{ c c c c c c c c c c }
      \toprule
       &  $x_{\mathbf{A}1}$ & $x_{\mathbf{B}1}$ &  $T_{1} \ (\text{K})$ & $x_{\mathbf{A}2}$ & $x_{\mathbf{B}2}$ & $T_{2} \ (\text{K})$ & $x_{\mathbf{A}3}$ & $x_{\mathbf{B}3}$ & $T_{3} \ (\text{K})$  \\
      \midrule
      $x_{0}$ &
        0.1405 & 0.6370 & 475.2 &
        0.1683 & 0.6240 & 484.2 &
        0.0545 & 0.5886 & 482.5 \\
      $x_{s1}$ &
        0.1436 & 0.6568 & 488.4 &
        0.1644 & 0.6375 & 481.1 &
        0.0510 & 0.6167 & 483.3 \\
      $x_{s2}$ &
        0.1287 & 0.6420 & 493.0 &
        0.1500 & 0.6230 & 485.3 &
        0.0450 & 0.5843 & 487.4 \\
      $x_{s3}$ &
        0.1757 & 0.6730 & 480.5 &
        0.1961 & 0.6533 & 472.7 &
        0.0650 & 0.6695 & 474.9 \\  
      $x_{s4}$ &
        0.1397 & 0.6534 & 489.7 &
        0.1608 & 0.6341 & 482.0 &
        0.0495 & 0.6086 & 484.1 \\
      $x_{s5}$ &
        0.1705 & 0.6715 & 481.5 &
        0.1904 & 0.6522 & 474.6 &
        0.0624 & 0.6623 & 476.4 \\

      \bottomrule
    \end{tabular}
\end{table*}

\begin{table}[t!]
  \renewcommand\arraystretch{1.25}
  \caption{Steady-state control inputs $u_{si}, i=1, \ldots, 5$, of reactor-separator process.}\label{table:sp:u}\vspace{5pt}
  \centering
    \begin{tabular}{ c c c c c c c c c c }
      \toprule
      & $Q_1~(\text{kJ/h})$ & $Q_2~(\text{kJ/h})$ & $Q_3~(\text{kJ/h})$  \\
      \midrule
      $u_{s1}$ &
      $2.87743 \times 10^6$ & $1.14562 \times 10^6$ & $2.95015 \times 10^6$ \\
      $u_{s2}$ &
      $2.98560 \times 10^6$ & $1.09962 \times 10^6$ & $2.97473 \times 10^6$ \\
      $u_{s3}$ &
      $2.93024 \times 10^6$ & $9.37622 \times 10^5$ & $2.93181 \times 10^6$ \\
      $u_{s4}$ &
      $2.97024 \times 10^6$ & $1.07729 \times 10^6$ & $2.93679 \times 10^6$ \\
      $u_{s5}$ &
      $2.83789 \times 10^6$ & $1.15186 \times 10^6$ & $2.84174 \times 10^6$ \\
      \bottomrule
    \end{tabular}
\end{table}

\subsection*{Results}

\begin{figure*}[t!]
    \centering
    \subfigure[Output trajectories based on Deep DeePC-I and constrained Deep DeePC-I.]{
    \label{fig:y-I}
    \includegraphics[width=1\textwidth]{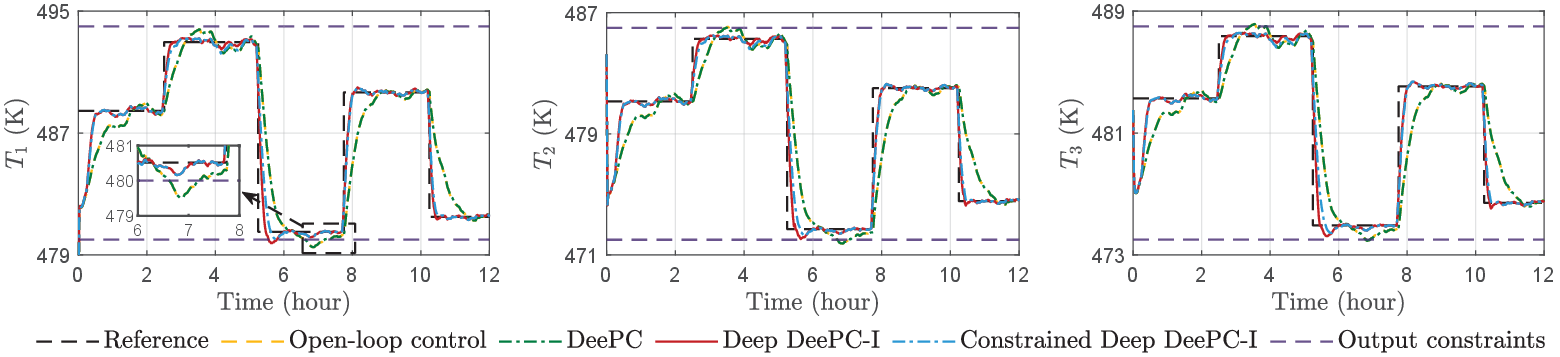}
    }

    \subfigure[Control input trajectories based on Deep DeePC-I and constrained Deep DeePC-I.]{
    \label{fig:u-I}
    \includegraphics[width=1\textwidth]{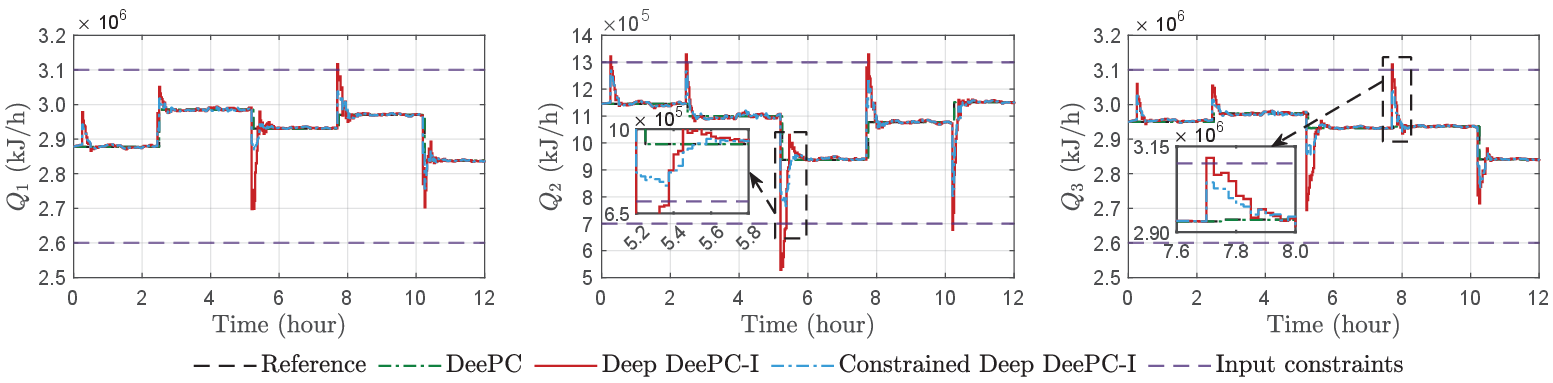}
    }
    \caption{Output and control input trajectories of reactor-separator process under designs with reference input.}
    \label{fig:yu-I}
\end{figure*}

First, we consider the case when the steady-state reference inputs corresponding to the set-points (reference outputs) are known. Two controllers are developed based on Deep DeePC-I and constrained Deep DeePC-I, respectively.

Figure \ref{fig:yu-I} presents the closed-loop output trajectories and input trajectories obtained based on the Deep DeePC-I, constrained Deep DeePC-I, and conventional DeePC. Figure \ref{fig:y-I} presents the closed-loop trajectories of the controlled output based on the three controllers. In addition, the output trajectories based on open-loop control, with inputs being made the same as the steady-state reference inputs, are provided for comparison. Figure \ref{fig:u-I} presents the input trajectories by different controllers. All three DeePC designs can track the set-points and maintain the operation level close to the set-points. The proposed Deep DeePC-I and constrained Deep DeePC-I outperform the conventional DeePC by converging faster to the set-points. As shown in Figure \ref{fig:u-I}, the control actions provided by Deep DeePC-I (shown in red lines) sometimes violate the 
input constraints. With the implementation of the proposed constraint handling scheme, constrained Deep DeePC-I (shown in blue dashed-dotted lines) ensures that the system constraints are satisfied throughout the entire operation.

\begin{figure*}[t!]
    \centering
    \subfigure[Output trajectories under the Deep DeePC-II and constrained Deep DeePC-II.]{
    \label{fig:y-II}
    \includegraphics[width=1\textwidth]{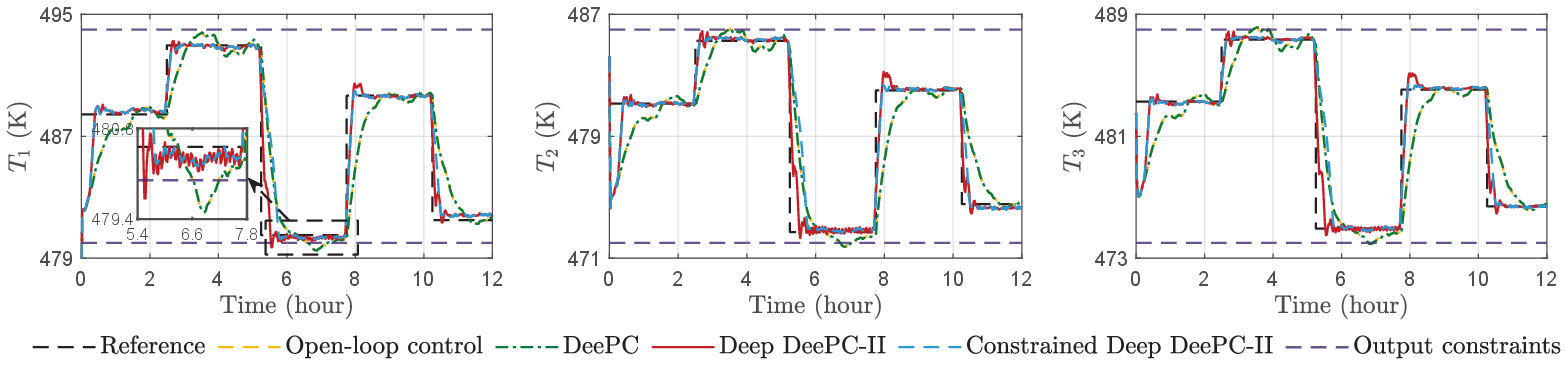}
    }

    \subfigure[Control input trajectories under the Deep DeePC-II and constrained Deep DeePC-II.]{
    \label{fig:u-II}
    \includegraphics[width=1\textwidth]{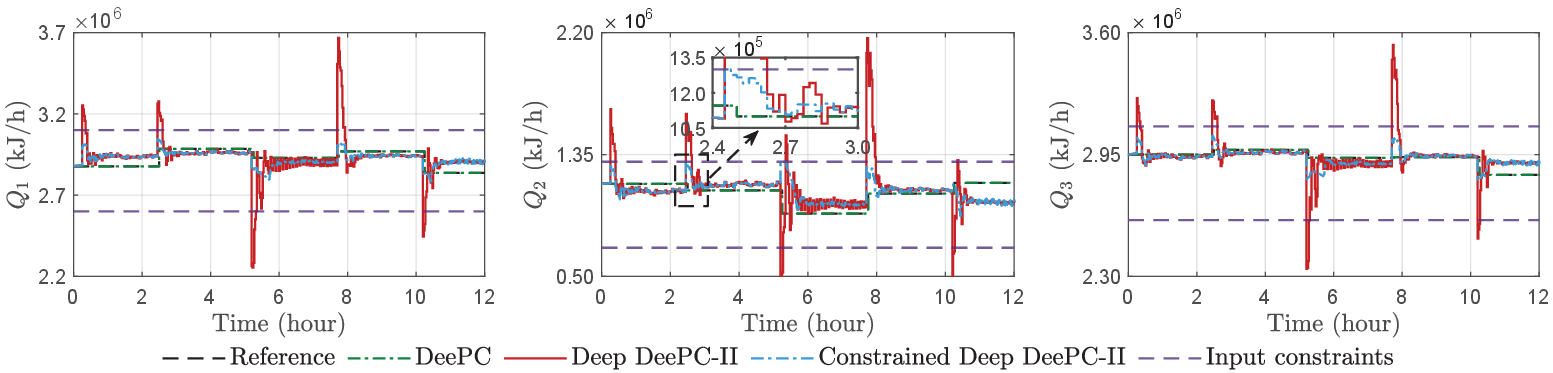}
    }
    \caption{Output and control input trajectories of reactor-separator process under designs without reference input.}
    \label{fig:yu-II}
\end{figure*}

Next, we consider the case when the steady-state reference inputs $\mathbf{u}^r$ are absent, and two controllers are developed based on Deep DeePC-II and constrained Deep DeePC-II, respectively. Figure \ref{fig:yu-II} presents the closed-loop results obtained based on different controllers. Specifically, Figure \ref{fig:y-II} presents the output trajectories of Deep DeePC-II, constrained Deep DeePC-II, and conventional DeePC; Figure \ref{fig:u-II} presents the corresponding control input trajectories. All the controllers are able to steer the process states toward the desired set-points and maintain the process operation close to the set-points. The proposed Deep DeePC designs demonstrate superior control performance compared to the conventional DeePC design in terms of faster convergence speed. As shown in Figure \ref{fig:u-II}, due to the absence of reference input, the control inputs provided by the Deep DeePC-II have a larger deviation from the steady-state reference inputs, as compared to the Deep DeePC-I design, which is designed and implemented with the steady-state reference inputs, as shown in Figure \ref{fig:u-I}. Although the Deep DeePC-II exhibits larger deviations from the steady-state reference inputs and significant input constraint violations, the proposed constraint handling scheme in constrained Deep DeePC-II is able to provide control actions that consistently satisfy the corresponding hard constraints on system inputs and outputs.

\begin{table}[t!]
  \renewcommand\arraystretch{1.25}
  \caption{Control performance comparison in terms of RMSEs for the case with steady-state reference inputs.}\label{table:rmse:I}\vspace{5pt}
  \centering\small
    \begin{tabular}{c c c c c }
      \toprule
      & Open-loop & DeePC & Deep DeePC-I & \thead{Constrained\\Deep DeePC-I} \\
      \midrule
      RMSE & 0.07537 & 0.07550 & 0.04194 & 0.04622 \\
      \bottomrule
    \end{tabular}
\end{table}

\begin{table}[t!]
  \renewcommand\arraystretch{1.25}
  \caption{Control performance comparison in terms of RMSEs for the case without steady-state reference inputs.}\label{table:rmse:II}\vspace{5pt}
  \centering\small
    \begin{tabular}{c c c c c }
      \toprule
      & Open-loop & DeePC & Deep DeePC-II & \thead{Constrained \\Deep DeePC-II} \\
      \midrule
      RMSE & 0.07537 & 0.07550 & 0.04309 & 0.06079 \\
      \bottomrule
    \end{tabular}
\end{table}

To quantitatively assess and compare the control performance, we compute the RMSE for each controller. The RMSEs are computed using scaled output values to mitigate the influence of multiple magnitudes of output variables that have different physical meanings. 
Table \ref{table:rmse:I} presents the RMSE results for Deep DeePC-I and constrained Deep DeePC-I, where steady-state reference inputs are used for controller training and online implementation. Both Deep DeePC designs outperform the conventional DeePC. The constrained Deep DeePC-I shows a slightly higher RMSE compared to Deep DeePC-I, due to the enforcement of system constraints. Table \ref{table:rmse:II} presents the RMSE results for Deep DeePC-II and constrained Deep DeePC-II, where steady-state reference inputs are absent. In this context, the proposed Deep DeePC-II and constrained Deep DeePC-II also provide improved performance compared to the conventional DeePC.

\begin{rmk}
Representative data-driven model predictive control methods typically utilize data information to develop a dynamic model of the system, and then formulate an online MPC optimization problem based on the data-driven model~\cite{patel2021model, tan2024robust, son2021application, han2023robust, xie2024data}. It is worth mentioning that these data-driven MPC methods typically require full-state information. In certain applications, measuring specific state variables in real-time with hardware sensors can be challenging or expensive. The Deep DeePC control scheme requires only control inputs and controlled outputs, eliminating the need to establish an explicit dynamic model. In such cases, the proposed approach can be more advantageous.
\end{rmk}

\subsection*{Comparisons of computation time}

We compare the computation time for conventional DeePC, Deep DeePC-I, and constrained Deep DeePC-I during online implementation. The average computation time per step across 100 evaluation trials is evaluated. Each evaluation trial starts with a different initial state and is subject to varying disturbances over a total of 500 sampling periods. The objective is to track the five set-points listed in Table \ref{table:sp:x}. The experiments are conducted on a computer equipped with an Intel$^{\circledR}$ Core$^{\text{TM}}$ i9-13900 CPU with 24 cores and 128 GB of random access memory (RAM).

Table \ref{table:time} presents the average computation time for different DeePC-based controllers. The proposed methods significantly reduce the average computation time compared to the conventional DeePC. Specifically, the average computation time for Deep DeePC-I, which does not involve online optimization, is reduced by 99.69\% compared to the conventional DeePC. The average computation time for constrained Deep DeePC-I is reduced by 93.15\% compared to the conventional DeePC.
The constrained Deep DeePC-I requires more computation time than Deep DeePC-I since it involves the event-based constraint handling scheme. This scheme is activated to manage hard constraints at an average event rate of 19.84\% across the 100 evaluation trials. The average computation time for a single execution of the constraint handling scheme in (\ref{opt:g}) is 0.01598 seconds, achieving a 67.23\% reduction in computation time compared to a single execution of conventional DeePC in (\ref{deepc_2}). This indicates that when the constraint handling scheme is activated, it solves a simpler optimization problem than the one associated with conventional DeePC.

\begin{table}[t!]
  \renewcommand\arraystretch{1.25}
  \caption{Comparison of computation time.}\label{table:time}\vspace{5pt}
  \centering\small
    \begin{tabular}{c c c c}
      \toprule
      & DeePC  &Deep DeePC-I & \thead{Constrained \\ Deep DeePC-I}  \\ 
      \midrule
         Time (s/step) & 0.04877 & 0.00015 & 0.00334 \\
      \bottomrule
    \end{tabular}
\end{table}

\begin{table}[t!]
  \renewcommand\arraystretch{1.25}
  \caption{Computation times of conventional DeePC with varying $T$ ($T_{ini}=N_p=10$).}\label{table:deepc:time}\vspace{5pt}
  \centering\small
    \begin{tabular}{c c c c}
      \toprule
      $T$ & 200  &400 & 600  \\ 
      \midrule
         Time (s/step) & 0.04877 & 0.14146 & 0.30272 \\
      \bottomrule
    \end{tabular}
\end{table}

We also assess how the parameters affect the computation time of conventional DeePC. Table \ref{table:deepc:time} presents the average computation times for DeePC with different values of $T$.  The results indicate that the computation time for the DeePC optimization problem increases rapidly as $T$ increases. This suggests that the complexity of the conventional DeePC optimization problem in (\ref{deepc_2}) is significantly influenced by the parameters $T$. To satisfy the persistently exciting condition described in Definition \ref{def1}, $T$ generally needs to be made sufficiently large, which leads to increased computation time for DeePC. In contrast, the proposed Deep DeePC methods are minimally affected by these parameters ($T$, $T_{ini}$, and $N_p$), since either no optimization or only low optimization needs to be solved during online implementation.

\section*{Conclusion}

In this paper, we proposed a computation-efficient, deep learning-enabled DeePC control approach for nonlinear processes, referred to as Deep DeePC. Within the proposed framework, a deep neural network was used to output the DeePC operator $g$ at each sampling instant, and the generated DeePC operator was then used to generate an optimal control input sequence. The parameters of this neural network can be trained using input and output data collected from open-loop process operations. During online implementation, the Deep DeePC framework leverages the trained neural network to directly output the DeePC operator. Therefore, online optimization required by conventional DeePC is bypassed. To further address constrained control problems, a constraint handling scheme was developed and integrated with the proposed Deep DeePC. 
The proposed approach was applied to both a gene regulatory network and a reactor-separator chemical process, good control performance was achieved in both case studies. The results demonstrated that this method can drive process operations toward desired set-points even when the reference inputs corresponding to those set-points are not provided. Additionally, the online computation time for executing the proposed Deep DeePC is significantly reduced compared to conventional DeePC. The constrained Deep DeePC can handle input and control constraints, while maintaining more efficient online computation as compared to conventional DeePC.

\section*{Acknowledgment}
This research is supported by the National Research Foundation, Singapore, and PUB, Singapore’s National Water Agency under its RIE2025 Urban Solutions and Sustainability (USS) (Water) Centre of Excellence (CoE) Programme, awarded to Nanyang Environment \& Water Research Institute (NEWRI), Nanyang Technological University, Singapore (NTU). 
This research is also supported by Ministry of Education, Singapore, under its Academic Research Fund Tier 1 (RS15/21 \& RG63/22), and Nanyang Technological University, Singapore (Start-Up Grant). 
Any opinions, findings and conclusions or recommendations expressed in this material are those of the author(s) and do not reflect the views of the National Research Foundation, Singapore and PUB, Singapore's National Water Agency.


\section*{Data availability statement}

The numerical data used to generate Figures \ref{fig:grn:yu}, \ref{fig:yu-I}, \ref{fig:yu-II}, and Tables \ref{table:grn:rmse}, \ref{table:rmse:I}, \ref{table:rmse:II} are provided in the Supplementary Material. The compressed file contains data obtained from the case studies on the gene regulatory network and the reactor-separator process. This encompasses simulated historical open-loop data for neural network training, maximum and minimum values of states used for scaling, and the simulated closed-loop trajectories based on different control approaches (i.e., our proposed Deep DeePC, conventional DeePC, and open-loop control). The data can be used to reproduce and generate the figures and tables presented in this paper. Specifically, the RMSEs in Tables \ref{table:grn:rmse}, \ref{table:rmse:I}, and \ref{table:rmse:II} can be obtained using the reference trajectories and the trajectories provided by different controllers. Additionally, the Hankel matrices and the training data for the neural network can be constructed using historical open-loop data, following the instructions provided in this paper. The source code for our methods is available at {\small{https://github.com/Zhang-Xuewen/Deep-DeePC}}.

\renewcommand{\refname}{Literature cited}

\end{document}